\documentclass[english]{article}
\usepackage[T1]{fontenc}
\usepackage[utf8]{inputenc}
\usepackage{babel}
\usepackage{float}
\usepackage{graphicx}
\usepackage{geometry}
\usepackage[]
 {hyperref}

\makeatletter

\providecommand{\tabularnewline}{\\}

\usepackage{authblk} 

\date{} 

\author{Tiago da Cruz} 
\author{Bernardo Tavares} 
\author{Francisco Belo}

 \affil{Granter.ai} 
 \affil{\texttt{tiagomartinsdacruz@gmail.com} \\ 
 	\texttt{bernardo.tavares@granter.ai}}

\usepackage{fancyvrb}

\makeatother

\begin{document}
\title{Ontology Learning and Knowledge Graph Construction: A Comparison of
Approaches and Their Impact on RAG Performance}
\maketitle
\begin{abstract}
Retrieval-Augmented Generation (RAG) systems combine Large Language
Models (LLMs) with external knowledge, and their performance depends
heavily on how that knowledge is represented. This study investigates
how different Knowledge Graph (KG) construction strategies influence
RAG performance. We compare a variety of approaches: standard vector-based
RAG, GraphRAG, and retrieval over KGs built from ontologies derived
either from relational databases or textual corpora. Results show
that ontology-guided KGs incorporating chunk information achieve competitive
performance with state-of-the-art frameworks, substantially outperforming
vector retrieval baselines. Moreover, the findings reveal that ontology-guided
KGs built from relational databases perform competitively to ones
built with ontologies extracted from text, with the benefit of offering
a dual advantage: they require a one-time-only ontology learning process,
substantially reducing LLM usage costs; and avoid the complexity of
ontology merging inherent to text-based approaches.
\end{abstract}

\section{Introduction}

Large Language Models have demonstrated strong generative and reasoning
abilities across a variety of domains, but their reliance on static
training data limits their access to unseen and domain-specific knowledge.
Retrieval-Augmented Generation (RAG) mitigates these limitations by
grounding model outputs in external sources, reducing hallucinations.
However, RAG performance depends on how that external knowledge is
structured and retrieved.

Traditional RAG systems rely on vector databases, which index text
embeddings for semantic similarity search. While efficient, this representation
ignores the relational structure between entities, possibly leading
to redundant retrieval and limited reasoning. Graph-based RAG approaches
address this by leveraging Knowledge Graphs to encode explicit relationships,
enabling structured and interpretable retrieval. Nonetheless, building
and maintaining high-quality KGs remains a non-trivial task \cite{app15073727,Hofer_2024,Ji_2022}.

Recent advances in ontology learning have opened new avenues for automating
KG construction. Ontologies formally capture domain concepts and relations,
serving as blueprints for graph construction. LLMs also enable ontology
extraction from both structured and unstructured data. Despite this
progress, the potential of ontology-guided KGs, particularly those
derived from relational databases, remains largely unexplored in the
context of RAG. This study addresses that gap by systematically comparing
how different KG construction strategies influence RAG performance. 

Using a real-world dataset (a grant application from Granter.ai) we
evaluate standard Vector RAG, GraphRAG (Microsoft Research’s GraphRAG
system), and retrieval over several KGs built from ontologies extracted
either from a relational database or directly from text. By controlling
for corpus, retrieval method, KG construction pipeline, and evaluation
framework, the analysis isolates the effect of ontology source and
structure on retrieval accuracy and reasoning quality.

The contribution of our work is two-fold:

First, we show that integrating textual chunks directly into the graph
structure, beyond traditional entity and relation-only Knowledge Graphs,
improves retrieval accuracy and generation quality by enriching context
and enabling deeper reasoning.

Second, we demonstrate that ontologies extracted from static relational
databases yield performance comparable to those derived from text
corpora, while being far more cost-efficient and easier to maintain.
Since database schemas tend to remain stable over time, the ontology
learning process needs to be performed only once, avoiding the high
computational cost of repeated LLM inference and the complexity of
ontology merging required in text-based approaches.

Together, these findings highlight practical strategies for building
scalable and interpretable Graph-based RAG systems.

Our code is publicly available: \href{https://github.com/tiagocrz/KGs_for_Vertical_AI}{https://github.com/tiagocrz/KGs\_for\_Vertical\_AI}

\section{Related work}

\subsection{RAG}

Retrieval-Augmented Generation has emerged as a widely adopted paradigm
to enhance Large Language Models with external knowledge, addressing
limitations such as hallucinations and lack of domain-specific awareness.
\cite{peng2024graphretrievalaugmentedgenerationsurvey}

In these pipelines, documents are embedded into a vector space, and
semantically similar chunks are retrieved to augment the LLM's prompt,
allowing for more accurate responses.

While effective for many applications, vector-based retrieval has
inherent limitations, such as neglecting relationships among the text
corpus and producing overly long contexts \cite{peng2024graphretrievalaugmentedgenerationsurvey,wang2023knowledgegraphpromptingmultidocument}.

To overcome these challenges, Graph-based RAG pipelines leverage structured
graphs to capture relations in the data. Recent surveys emphasize
that using graphs improves interpretability, reduces redundancy in
retrieval, and supports tasks requiring multi-hop reasoning and global
summarisation, areas where vector-based methods struggle \cite{peng2024graphretrievalaugmentedgenerationsurvey,zhong2023comprehensivesurveyautomaticknowledge,app15073727}
.

Hybrid approaches that combine vector retrieval and graph-based retrieval
have also become more prominent. For example, HybridRAG \cite{sarmah2024hybridragintegratingknowledgegraphs}
integrates both Knowledge Graphs and vector databases for document
analysis, showing superior retrieval accuracy and answer quality compared
to using either a vector-based, or a graph-based approach alone.

Empirical studies also highlight the importance of evaluation. For
instance, Han et al. \cite{han2025ragvsgraphragsystematic} compare
these methods directly, showing that graph-based RAG outperforms in
interpretability and reasoning, whereas vector RAG remains competitive
in detailed single-hop queries. Similarly, Myers et al. \cite{myers2025talkinggdeltknowledgegraphs}
explore multiple Knowledge Graph construction options for RAG, concluding
that hybrid methods may be the optimal approach.

\subsection{Knowledge Graphs}

Knowledge Graphs (KGs) provide a structured representation of knowledge,
typically expressed as entities, relationships, and semantic descriptions.
By organizing information in this way, KGs enable semantic querying,
reasoning, and integration across heterogeneous sources \cite{Ji_2022,nayyeri2025retrievalaugmentedgenerationontologiesrelational}.
They have become a fundamental data structure for capturing domain
knowledge, supporting applications such as semantic search, recommendation
systems, and question answering \cite{app15073727}.

The KG construction process can be automated using rule-based, statistical,
or LLM-driven methods, each with distinct trade-offs between precision
and scalability \cite{zhong2023comprehensivesurveyautomaticknowledge}.
Choi and Jung \cite{app15073727} summarize KG construction into three
phases: extraction (from structured, semi-structured, or unstructured
data), learning (embedding and inference techniques), and evaluation
(measures of accuracy, coverage, and consistency). 

Despite recent progress, several issues persist. Research consistently
points to two broad challenges: construction complexity, as high-quality
KGs require integrating diverse data sources, and maintenance, as
KGs must adapt to new data \cite{app15073727,zhong2023comprehensivesurveyautomaticknowledge}.

Studies such as Hofer et al. \cite{Hofer_2024} also stress that some
examples of current bottlenecks, specifically to Knowledge Graph Construction,
relate to ontology alignment and schema evolution. Addressing these
challenges has motivated the rise of ontology-grounded approaches,
which play a key role in domains that demand high interpretability
and structured reasoning \cite{myers2025talkinggdeltknowledgegraphs}.
These ontology-guided KGs serve as the foundation for the comparative
analysis conducted in this study.

\subsection{Ontology Learning}

Ontologies are formal representations of knowledge that define concepts,
relationships, and constraints within a specific domain. When used
to guide knowledge graph construction, they enable more consistent,
hierarchical representations of knowledge and support complex tasks
such as multi-hop reasoning, disambiguation, and context-aware retrieval
\cite{peng2024graphretrievalaugmentedgenerationsurvey,myers2025talkinggdeltknowledgegraphs}.

Recently, LLMs have shown significant promise in automating ontology
learning from both structured and unstructured data. Methods such
as the ones proposed by Nayyeri et al. \cite{nayyeri2025retrievalaugmentedgenerationontologiesrelational}
and Bakker et al. \cite{Bakker2024} use prompt-based approaches to
extract ontology elements directly from relational databases or text,
bypassing traditional manual and rule-based systems. 

Other studies, such as Chepurova et al. \cite{chepurova-etal-2024-prompt}
and Doumanas et al. \cite{app15042146}, demonstrate that integrating
ontology verification or fine-tuning LLMs on ontology-engineering
corpora improves precision in domain alignment, allowing for ``more
intelligent and automated systems in ontology management'' \cite{app15042146}.
Even so, challenges remain in ensuring logical consistency, mitigating
hallucinations, and aligning generated ontologies with established
vocabularies and schema standards \cite{Bakker2024,app15073727}.

In our study, ontologies serve to guide the construction of the KGs
that will be at the basis of the RAG systems tested.

\section{Methodology}

This section describes the experimental pipeline implemented to compare
different KG construction methods and their integration with RAG.
We evaluate six approaches: a baseline vector-based RAG, GraphRAG,
and retrieval over four different KGs.

\begin{figure}[H]
\centering
\includegraphics[scale=0.35]{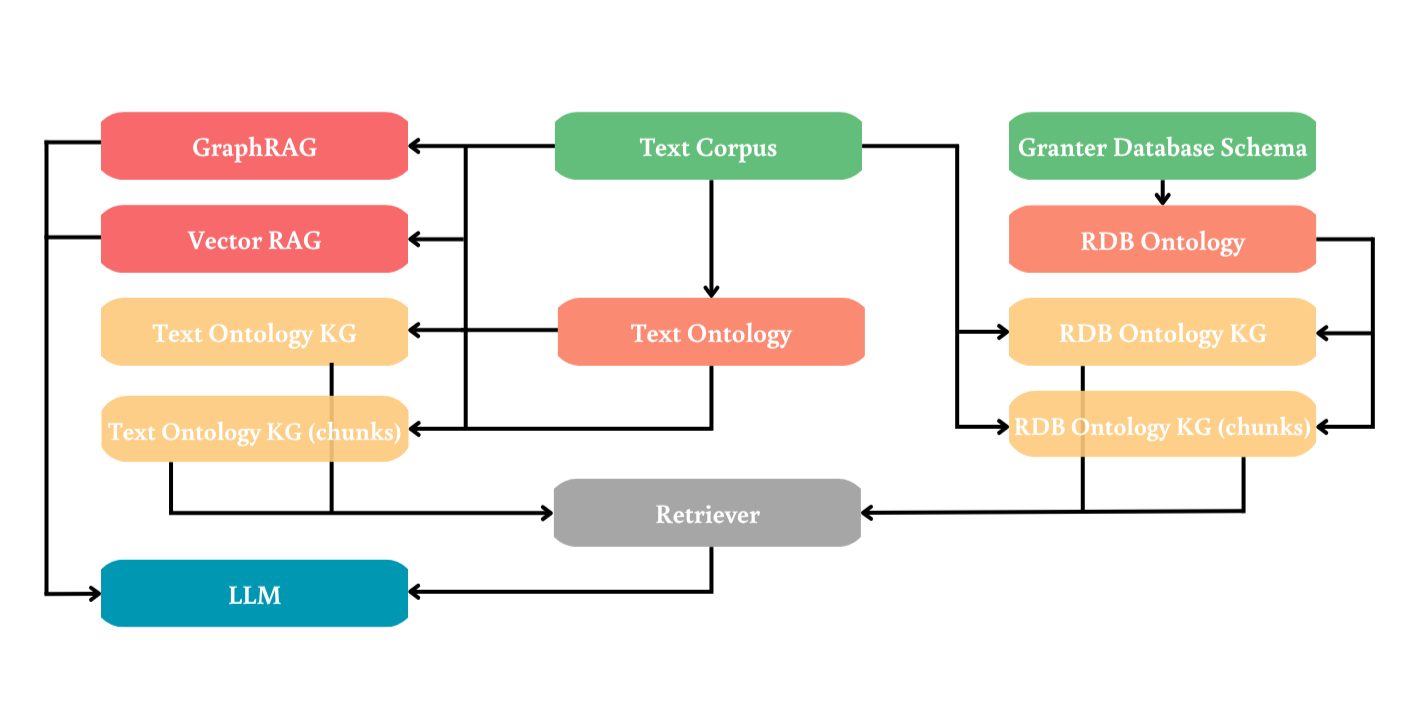}

\caption{{\small Experimental setup comparing Vector RAG, GraphRAG, and ontology-guided
Knowledge Graphs.}}

\end{figure}

\subsection{RAG baselines}

\subsubsection{Vector RAG}

The first baseline follows the conventional vector-based RAG paradigm. 

Documents in the corpus were split into overlapping text chunks using
LangChain’s RecursiveCharacterTextSplitter, and were subsequently
embedded using Ollama’s nomic-embed-text:v1.5 model. The embeddings
were indexed with FAISS, and retrieval was carried out using cosine
similarity to identify the nearest neighbours.

This configuration represents the canonical pipeline for VectorRAG
and serves as a reference point for assessing the added value of graph-enhanced
approaches.

\subsubsection{GraphRAG}

We implemented Microsoft's GraphRAG framework \cite{edge2025localglobalgraphrag},
which extends RAG with graph-based summarisation and retrieval. GraphRAG
constructs an entity graph from the corpus and leverages community
structures to enable retrieval of semantically relevant subgraphs.

The implementation relied on the official Python package (graphrag)
and was run with its default settings, including the local search
method for retrieval. This provided an out-of-the-box reference system
without manual tuning.

\subsection{Ontology-Grounded KGs}

\subsubsection{RDB Ontology Learning}

The first ontology-driven Knowledge Graph (KG) construction method
builds upon RIGOR (Retrieval-Augmented Iterative Generation of Relational
Database Ontologies) \cite{nayyeri2025retrievalaugmentedgenerationontologiesrelational},
which generates ontologies directly from relational database (RDB)
schemas.

We implemented an adapted version of RIGOR, introducing the following
changes:
\begin{itemize}
\item Instead of leveraging multiple external ontology repositories, we
restricted the external reference ontology to DINGO \cite{chialva2020dingoontologyprojectsgrants},
an ontology that provides context relative to projects, funding, actors,
and funding policies. In our implementation, DINGO was manually converted
to OWL 2 Manchester syntax for more accessible lexical extraction
of the ontological constraints;
\item In contrast to the original RIGOR pipeline, we did not rely on the
lexical view of the RDB. Instead, the input to the LLM included directly
the DDL statements of the tables with their columns and primary/foreign
key annotations;
\item No table-level documentation was available beyond general schema descriptions,
so we did not include any in the process;
\item Both the delta ontologies and the resultant ontology were generated
in Turtle (TTL) format, which facilitated iterative merging via the
RDFLib Python library;
\item LLM inference was ran with temperature = 0, prioritising strict instruction-following;
\item The prompting strategy was adjusted to enforce strict naming conventions,
ensuring that semantically identical elements were not redundantly
generated under different names. In addition, we instructed the LLM
to produce labels and comments for all ontology elements to support
interpretability;
\end{itemize}
These modifications yielded a consistent ontology extraction process
aligned with the database schema while minimising redundancy and noise.

\begin{figure}[H]
\centering
\includegraphics[scale=0.45]{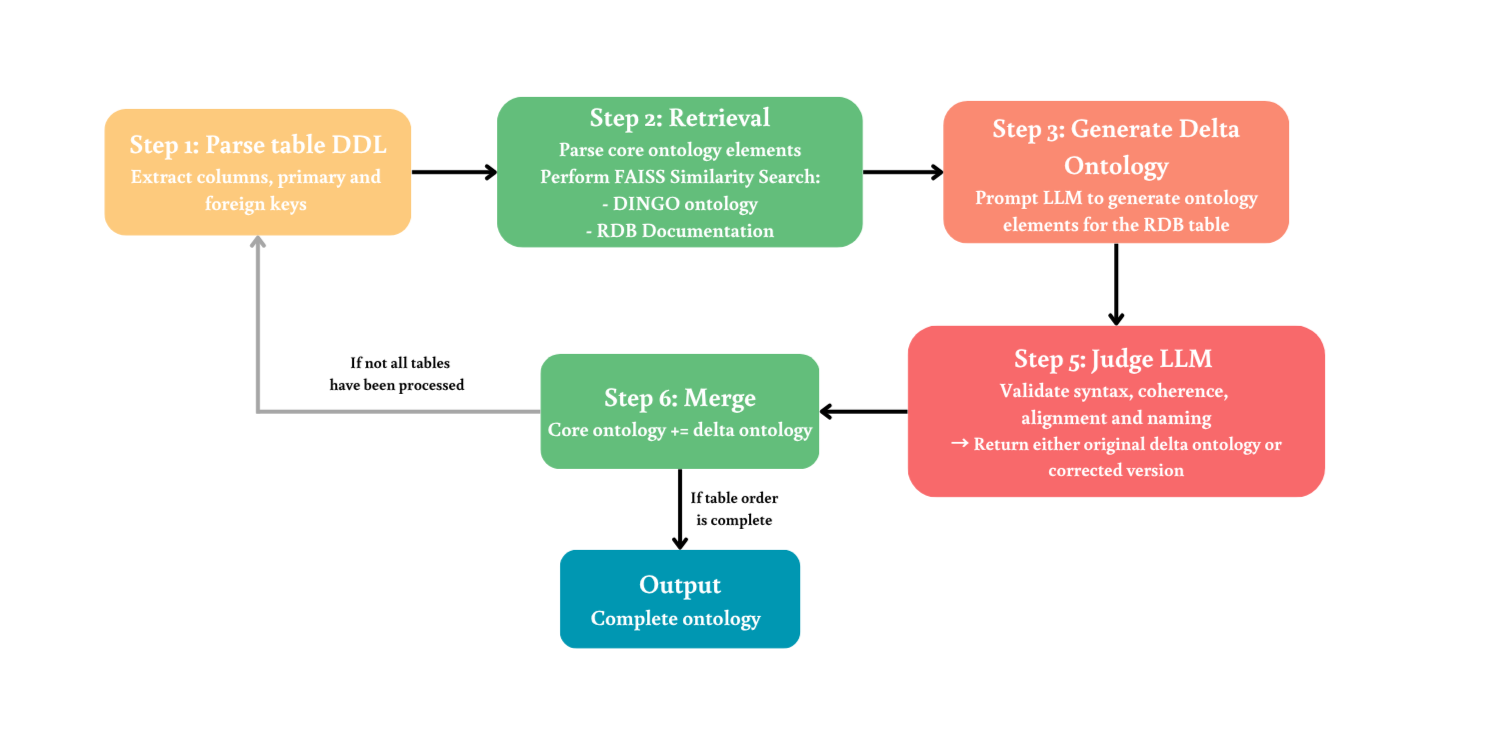}

\caption{Overview of the RIGOR Pipeline implemented.}

\end{figure}

\subsubsection{Text Ontology Learning}

The second ontology-driven method derives ontologies directly from
the text corpus, following the methodology proposed by Bakker et al.\cite{Bakker2024}
in their third extraction approach described, with a few minor adjustments:
\begin{itemize}
\item Prompts were adapted to enforce strict naming conventions and improve
parsing reliability;
\item For each sentence, the system attempted to generate valid Turtle syntax.
If parsing failed, a secondary LLM check was triggered to repair the
syntax and output a corrected version.
\end{itemize}
This approach enabled the automatic induction of a domain-specific
ontology directly from unstructured text.

\subsubsection{KG Construction Framework}

To integrate the ontologies with the RAG pipeline, we developed a
KG construction framework built on the LLMGraphTransformer module
from the LangChain Experimental library. The pipeline extracted rules
and constraints from the ontologies and used the text corpus to instantiate
them as graph structures. 

Key implementation details included:
\begin{itemize}
\item A pipeline to extract ontology constraints in the form of lists (of
allowed classes and relations);
\item Testing variations with and without chunk information, to examine
how information granularity affects graph formation and downstream
retrieval;
\item Storing outputs in CSV files, ensuring compatibility with the retrieval
process.
\end{itemize}
This framework provided a uniform interface for graph-based retrieval.

\subsection{Retriever}

For retrieval over the constructed graphs, we implemented a custom
retriever inspired by G-Retriever \cite{he2024gretrieverretrievalaugmentedgenerationtextual}.
Unlike the original, which employs graph neural networks (GNNs) to
model the structure of the retrieved subgraph, our version relied
on a simplified heuristic-based approach without GNNs.

Particularly, the retriever employed the following steps:
\begin{enumerate}
\item The knowledge graph is ingested from CSV format and split into nodes
(with textual attributes) and edges (with labeled relations).
\item Each node’s textual description is embedded using a SentenceTransformer
model, producing a semantic vector space for similarity matching.
The user query is then embedded in the same space, and the retriever
computes cosine similarity to identify the top-k most relevant nodes.
\item To incorporate relational context, the pipeline constructs an undirected
edge set and assigns uniform costs.
\item Nodes ranked highly in the similarity search are assigned prizes proportional
to both their similarity score and rank.
\item Using the Prize-Collecting Steiner Tree (PCST) optimization, a connected
subgraph of relevant nodes and edges is extracted, maximizing the
number of nodes and minimizing the number of edges retrieved.
\item The selected subgraph is converted into a human-readable context string,
where nodes are labeled with their attributes and edges are expressed
as triples. This string forms the context passed to the LLM.
\end{enumerate}
The aim behind this design was to:
\begin{itemize}
\item Maintain consistency across graph variants, ensuring that differences
in performance reflect the quality of the underlying KGs rather than
the complexity of the retriever;
\item Preserve interpretability and reproducibility while minimising potential
confounding factors.
\end{itemize}
While prior work suggests fine-tuning may improve retrieval \cite{myers2025talkinggdeltknowledgegraphs},
our focus is on comparing KG quality rather than maximizing retriever
performance. By employing this structured yet simple retrieval process,
we guarantee that differences in performance arise primarily from
the quality of the underlying knowledge graph rather than from complex
retrieval heuristics.

\begin{figure}[H]
\centering
\includegraphics[scale=0.4]{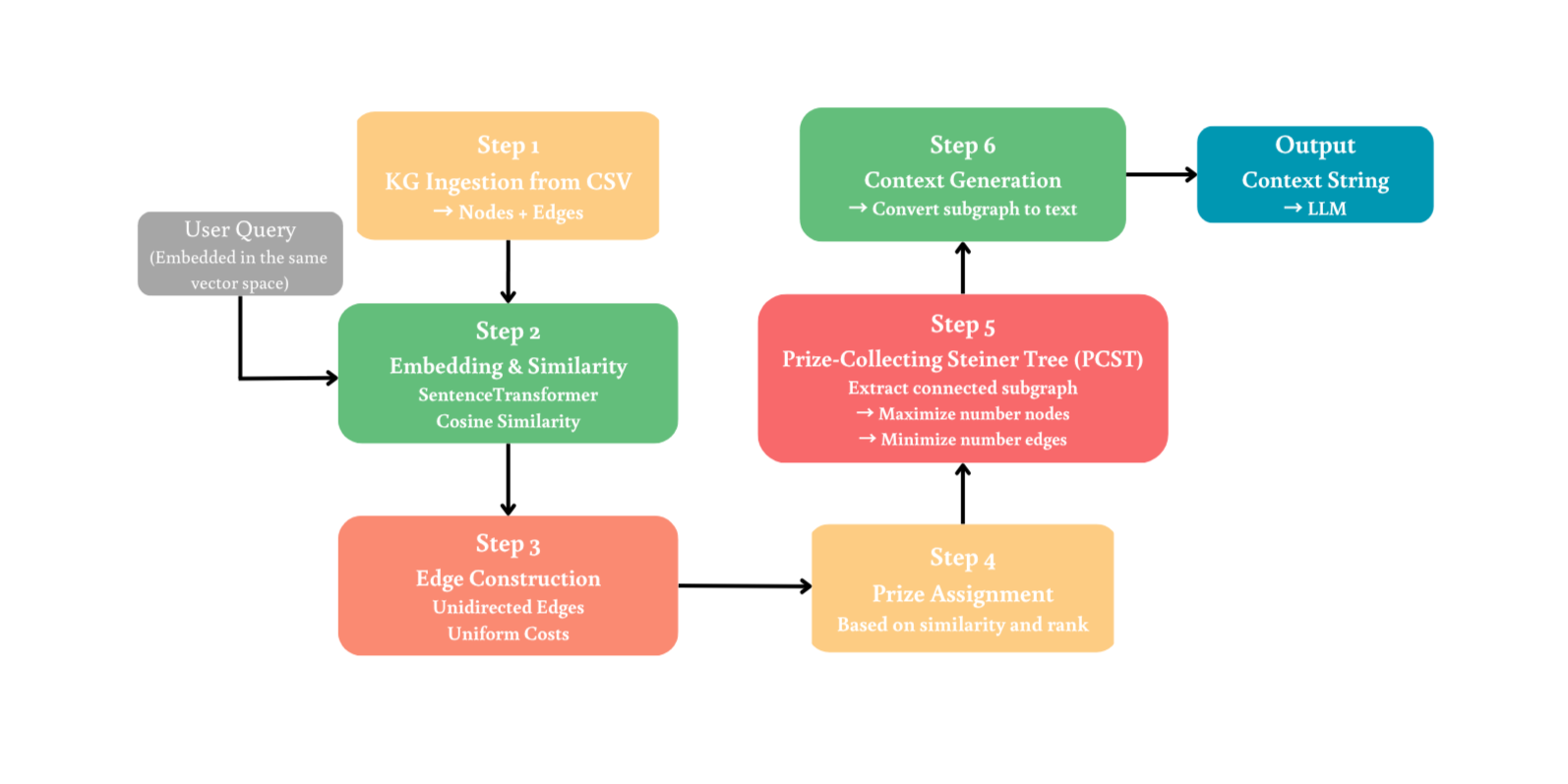}

\caption{Custom Retriever Pipeline.}

\end{figure}

\section{Experimental Setup}

\subsection{Corpus Description}

The experiments were conducted on a real grant application submitted
by Granter.ai to the ``Vouchers for Startups -- New Digital/Technological
Products'' call in Portugal, which was successfully approved. The
original document, written in Portuguese, was translated into English
and minimally anonymized.

This document was selected because it provides a representative real-world
example of a document that contains both company-level context (mission,
business model, technical and managerial capacity) and opportunity-specific
content (objectives, typology justification, innovation potential,
and impact).

This structure provides a rich and coherent dataset for evaluating
knowledge retrieval and reasoning performance, as it contains detailed
factual and contextual information relevant to multiple dimensions
of Granter’s operations.

\subsection{Question Set and Ground Truth}

To evaluate each method, a set of 20 domain-specific questions was
manually created to reflect the type of information expected to be
retrieved by real tasks in Granter's workflow.

These questions span five main categories:
\begin{enumerate}
\item Corporate and contextual information (e.g., “What is Granter.ai’s
main business activity?”)
\item Technical and methodological aspects (e.g., “Which AI methodologies
does Granter.ai use?”)
\item Strategic and managerial details (e.g., “Which team member is responsible
for the international expansion of the AI Agent?”)
\item Financial and operational metrics (e.g., “How much funding did Granter.ai
raise?”)
\item Scalability and impact (e.g., “What makes Granter.ai’s solution scalable
to different sectors?”) 
\end{enumerate}
Each question was paired with a reference answer extracted from the
application document, forming a ground truth dataset.

This ground truth was used for both qualitative comparison (completeness
and factuality) and quantitative evaluation.

\subsection{Evaluation Procedure}

Each retrieval method was evaluated on the same set of 20 questions
described in Section 4.2.

To assess response quality, a manual categorical evaluation framework
was adopted.

For every generated answer, the output was compared against the ground
truth and classified into one of four categories:
\begin{itemize}
\item Correct: the answer fully matches the ground truth in both factual
content and completeness;
\item Incomplete: the answer contains partial but accurate information,
omitting relevant details present in the ground truth;
\item False/Wrong: the answer includes incorrect or misleading information
not supported by the corpus;
\item “I don’t know”: the model explicitly indicates a lack of sufficient
information or refuses to answer.
\end{itemize}
This qualitative evaluation accurately captures the performance characteristics
of RAG systems in knowledge-intensive tasks, especially when outputs
may be partially correct or abstain from guessing.

The results of the tested retrieval approaches are reported in Section
5.

\section{Results}

\subsection{Overview}

The results of the six configurations are summarised in Figure 4.

The table reports the number of answers in each evaluation category
(Correct, Incomplete, False/Wrong, and “I don’t know”).

\begin{figure}[H]
\centering
\includegraphics[scale=0.3]{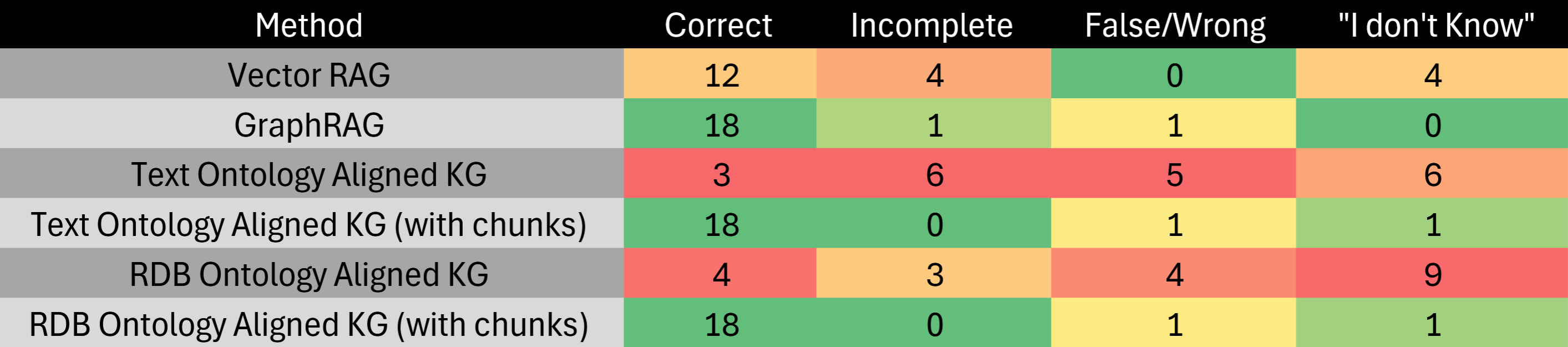}

\caption{Overview of results}
\end{figure}

\subsection{Key Insights and Comparative Evaluation}

The GraphRAG methodology and both ontology-guided Knowledge Graphs
that included chunk information achieved the highest accuracy, correctly
answering 18 out of 20 questions (90\%) each, with only one incorrect
and one “I don’t know” response. This shows that retrieval over an
ontology-constrained KG can match the performance of a state-of-the-art
framework when text segments are integrated directly into the graph
structure.

In contrast, the ontology-based graphs without chunk information performed
substantially worse. The Text Ontology KG achieved only 15\% accuracy
(3/20) and 30\% incomplete answers, while the RDB Ontology KG reached
20\% accuracy (4/20) and 45\% unanswered questions. These results
indicate that adding textual chunks to graph nodes significantly improves
factual grounding and completeness, without adding implementation
complexity.

The baseline Vector RAG achieved 60\% accuracy (12/20) and 20\% incomplete
answers, confirming that traditional embedding-based retrieval remains
reliable but lacks the relational reasoning capabilities of graph-based
methods.

Overall, chunk integration consistently increased answer accuracy
and completeness across all configurations. The combination of symbolic
structure and contextual text segments proved critical for accurate
reasoning, highlighting the value of ontology-guided, hybrid graph
representations in RAG systems.

\section*{Conclusion}

This study compared various Knowledge Graph construction strategies
and examined their impact on RAG performance. The results suggest
that the way KGs are constructed significantly influences both retrieval
accuracy and generative quality. Ontology-grounded approaches, particularly
those incorporating textual chunk information, achieved the highest
performance, maintaining high interpretability and minimal hallucination.

A key finding is that aligning a Knowledge Graph with an ontology
extracted from a static relational database yields performance comparable
to ontologies derived from dynamic text corpora. This approach presents
two major practical advantages:
\begin{itemize}
\item First, ontology learning from relational databases only needs to be
performed once, as relational schemas tend to remain stable over time,
which significantly reduces computational costs associated with repeated
LLM inference.
\item Second, this approach eliminates the need for complex ontology-merging
frameworks. In text-based ontology learning, every newly added document
can introduce redundant or conflicting entities, requiring sophisticated
merging and alignment pipelines. In contrast, database-derived ontologies
offer a stable schema that simplifies the maintenance and updating
of domain ontologies.
\end{itemize}
Together, these findings suggest that ontology extraction from relational
databases provides a scalable, cost-efficient solution for integrating
symbolic structure into RAG pipelines, particularly for industries
with stable and structured data sources.

It should be noted that these conclusions are drawn from an empirical
study with a limited dataset. While promising, the results should
be further validated across different domains and larger datasets
to assess the generalizability of the proposed approach. We leave
this as an avenue for future work.

\section*{Future Work }

While this study provides valuable insights into ontology-guided Knowledge
Graphs and their integration into RAG systems, several areas remain
for future exploration. To name the ones we consider to be the most
relevant:

\emph{Testing Text Ontology vs. RDB Ontology with Larger Corpora:}
Future research could extend this study by testing the performance
of text-derived ontologies versus relational database-derived ontologies
on a larger, more diverse corpus. This would allow for a more precise
quantification of the cost advantages of RDB-based ontologies, particularly
in terms of computational costs, while ensuring that performance is
maintained across more complex datasets.

\emph{Scalability Across Various Industries:} Future work could explore
whether, in contexts where text data is more dynamic and continuously
growing, the performance between RDB ontology-based and text ontology-based
Knowledge Graphs remains similar. This would provide a clearer understanding
of how stable and dynamic data environments affect retrieval performance
in different industries.

\emph{Ontology Learning from Relational Databases for RAG Systems:}
A particularly promising avenue for future work lies in deepening
the focus on ontology learning from relational databases specifically
for RAG systems. While this study demonstrated its efficiency and
stability, further work should investigate how to optimize and generalize
this process for diverse industrial settings, reinforcing its role
as a foundation for scalable and interpretable RAG pipelines.

\bibliographystyle{plain}
\bibliography{references}

\section*{Appendix}

\begin{table}[H]
\centering
\begin{tabular}{|c|c|}
\hline 
Method & File\tabularnewline
\hline 
\hline 
GraphRAG & \href{https://github.com/tiagocrz/KGs_for_Vertical_AI/blob/main/results/answers/graphrag/graphrag_results_localmethod.txt}{graphrag\_results\_localmethod.txt}\tabularnewline
\hline 
Vector RAG & \href{https://github.com/tiagocrz/KGs_for_Vertical_AI/blob/main/results/answers/vector_rag/rag_results.txt}{rag\_results.txt}\tabularnewline
\hline 
Text Ontology Aligned KG & \href{https://github.com/tiagocrz/KGs_for_Vertical_AI/blob/main/results/answers/txt_ontology/txt_results.txt}{txt\_results.txt}\tabularnewline
\hline 
Text Ontology Aligned KG (with chunks) & \href{https://github.com/tiagocrz/KGs_for_Vertical_AI/blob/main/results/answers/txt_ontology/txt_chunks_results.txt}{txt\_chunks\_results.txt}\tabularnewline
\hline 
RDB Ontology Aligned KG & \href{https://github.com/tiagocrz/KGs_for_Vertical_AI/blob/main/results/answers/rdb_ontology/rdb_results.txt}{rdb\_results.txt}\tabularnewline
\hline 
RDB Ontology Aligned KG (with chunks) & \href{https://github.com/tiagocrz/KGs_for_Vertical_AI/blob/main/results/answers/rdb_ontology/rdb_chunks_results.txt}{rdb\_chunks\_results.txt}\tabularnewline
\hline 
Ground Truth & \href{https://github.com/tiagocrz/KGs_for_Vertical_AI/blob/main/data/test_queries_groundtruth.txt}{test\_queries\_groundtruth.txt}\tabularnewline
\hline 
\end{tabular}

\caption{Answers produced by each method}

\end{table}

\begin{figure}[H]
\centering
\includegraphics[scale=0.2]{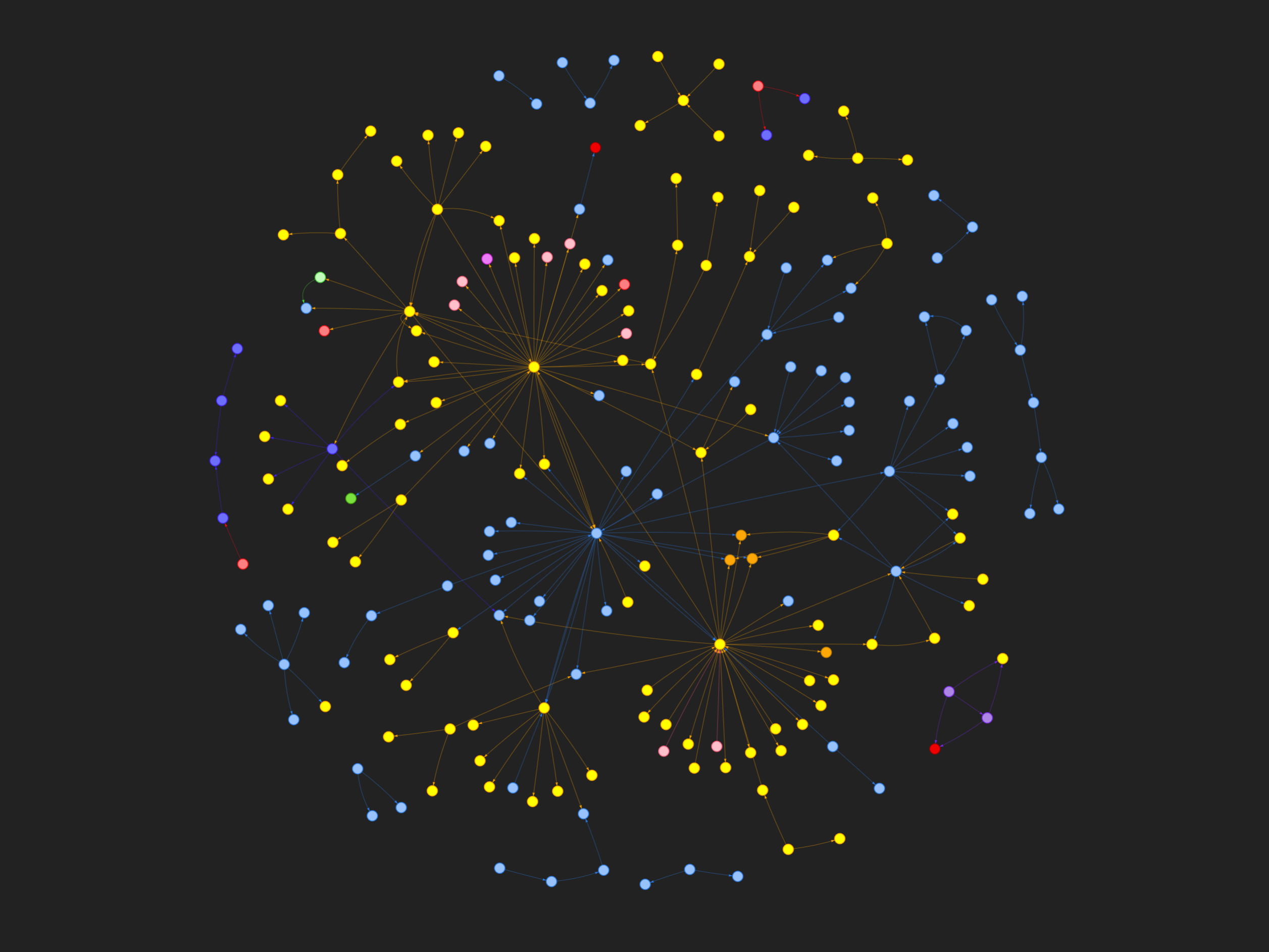}

\caption{RDB ontology aligned KG}

\end{figure}

\begin{figure}[H]
\centering
\includegraphics[scale=0.2]{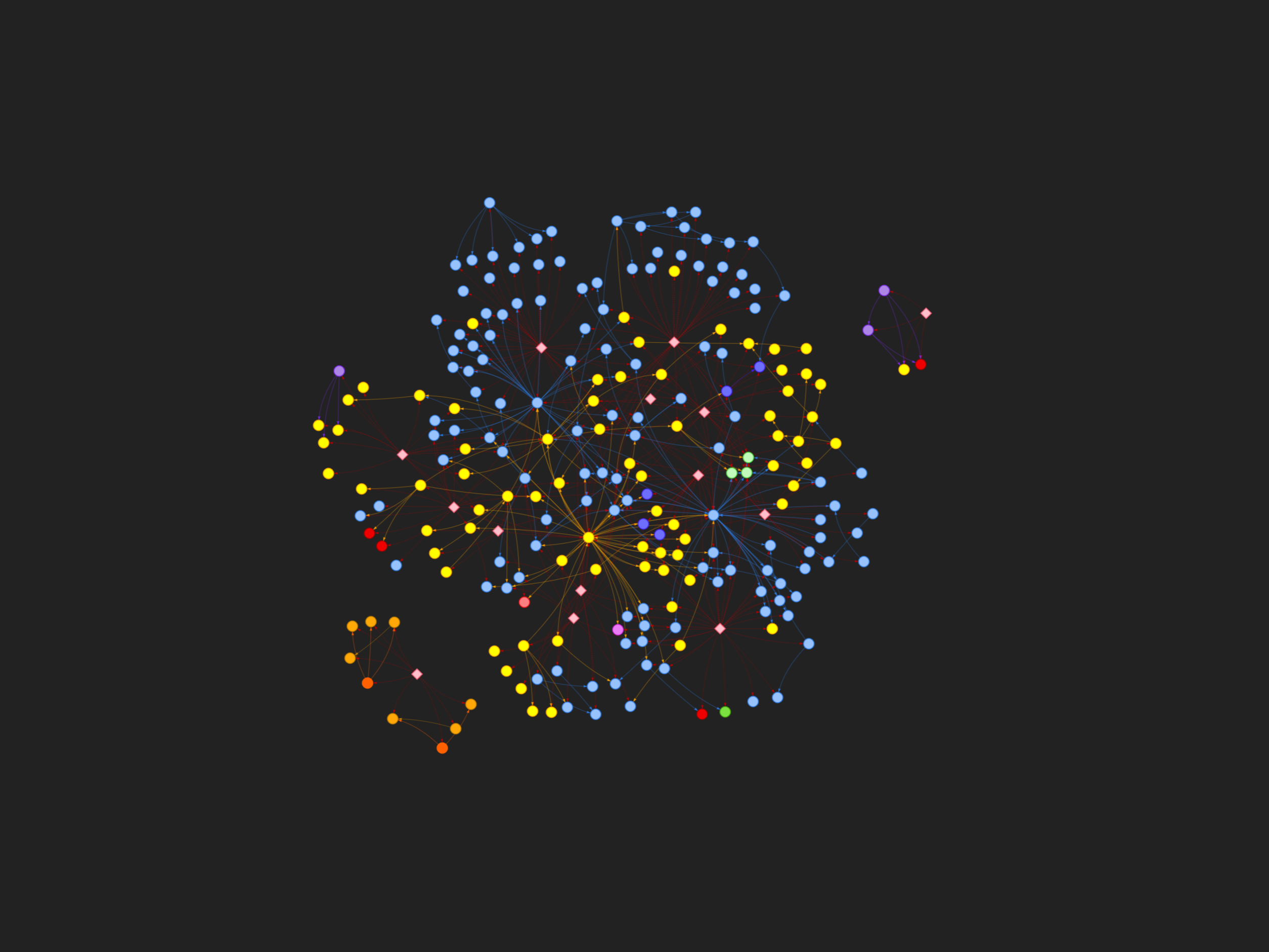}

\caption{RDB ontology aligned KG (with chunks)}

\end{figure}

\begin{figure}[H]
\centering
\includegraphics[scale=0.2]{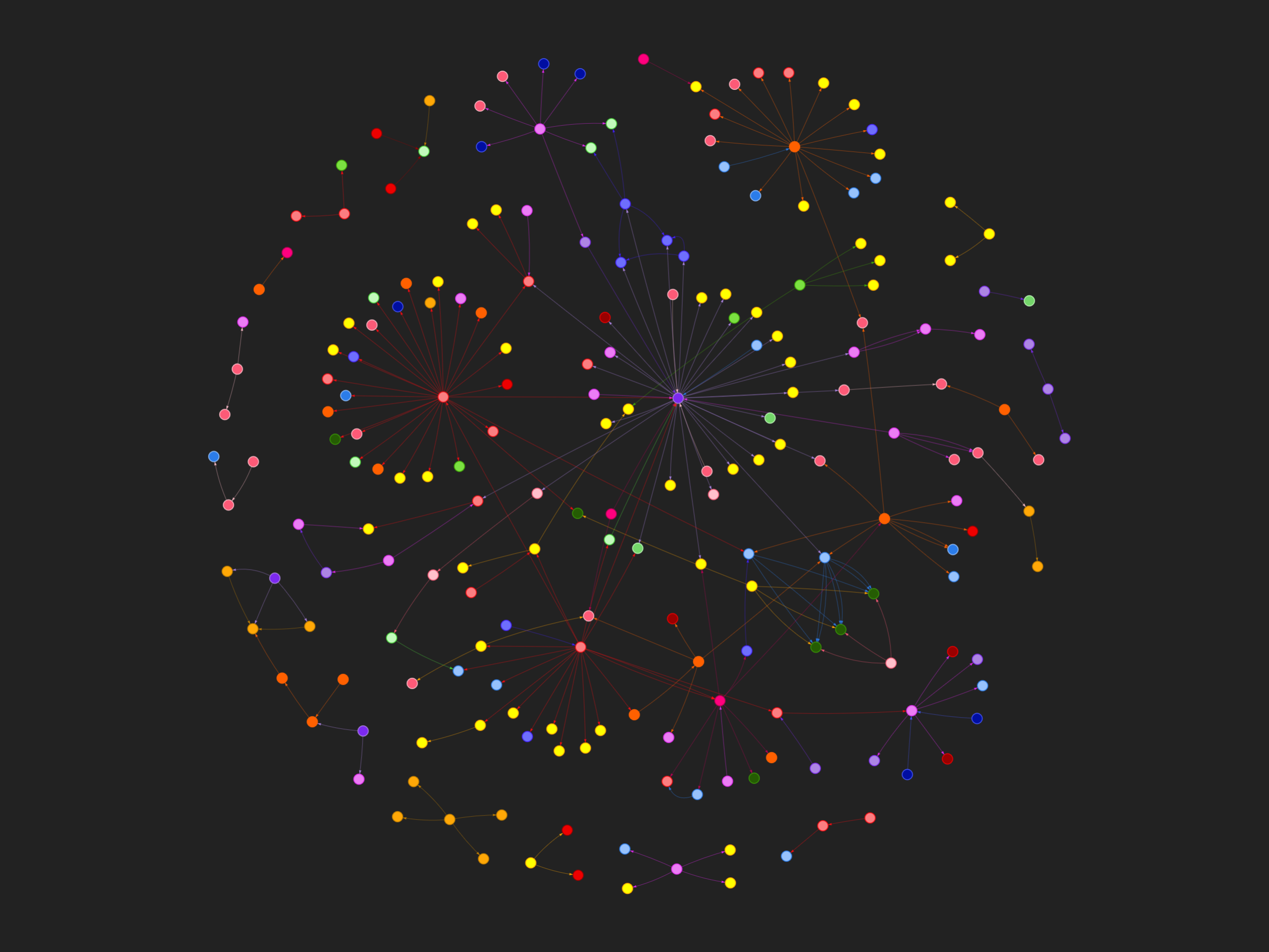}

\caption{Text ontology aligned KG}

\end{figure}

\begin{figure}[H]
\centering
\includegraphics[scale=0.2]{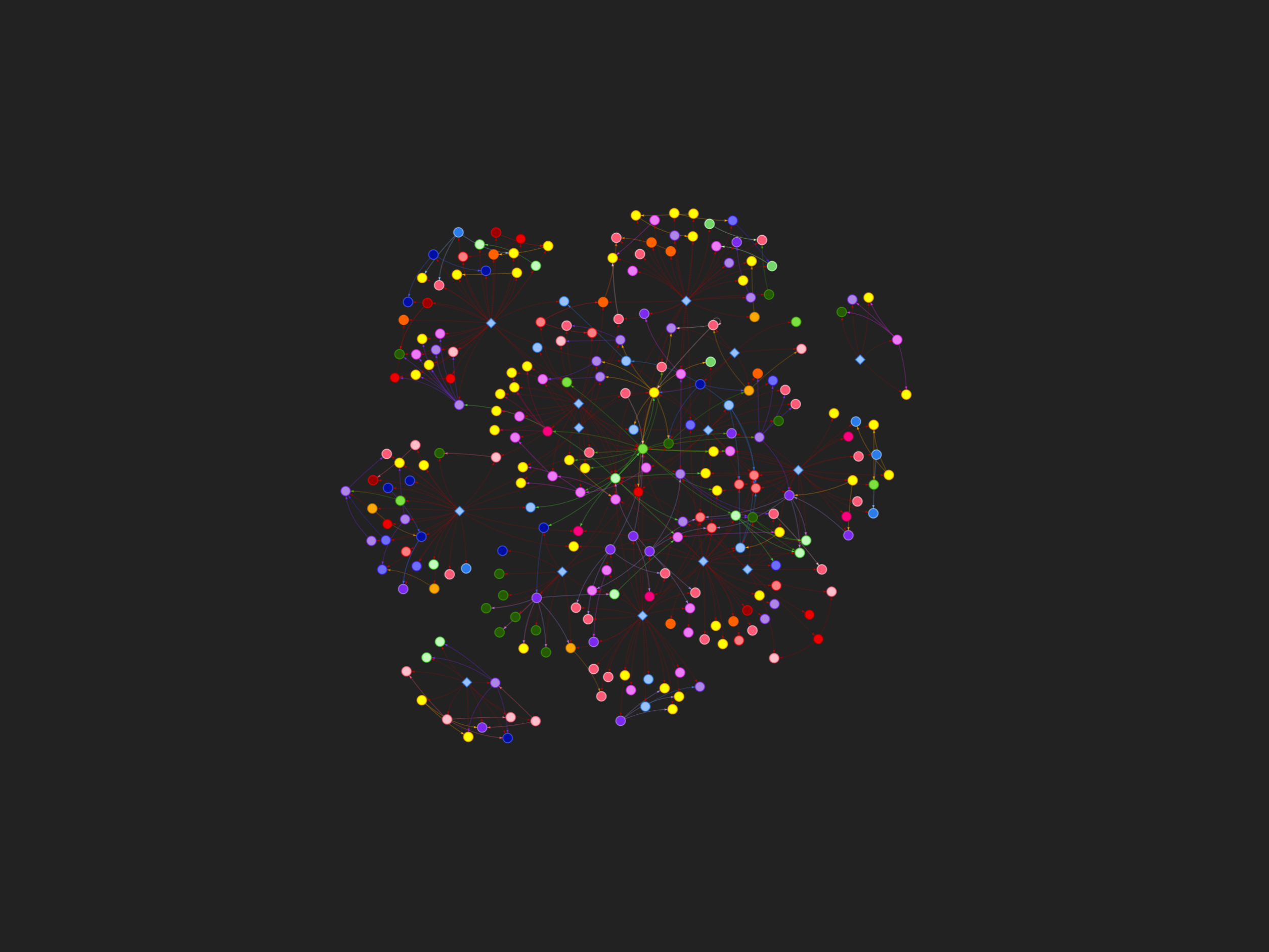}

\caption{Text ontology aligned KG (with chunks)}

\end{figure}

\end{document}